\documentclass{PoS}

\PoS{PoS(BDMH2004)034}

\usepackage{epsfig}
 
\title{Kinematics in Hickson Compact Group 90}

\ShortTitle{Kinematics in Hickson Compact Group 90}

\author{\speaker{Leonardo Casta\~neda-Colorado}\\
        Sternwarte der Universität Bonn, Germany\\
        E-mail: \email{leonardo@astro.uni-bonn.de}}

\author{Michael Hilker\\
        Sternwarte der Universität Bonn, Germany\\
        E-mail: \email{mhilker@astro.uni-bonn.de}}

\abstract{Using long-slit spectra observed at the VLT (ESO, Paranal) in May 
2002, we studied the stellar kinematics in the core of the Hickson Compact 
Group 90. The line of sight velocity distribution (LOSVD), the velocity 
dispersion profiles and the Hermite moments $h_3$, $h_4$ along two slit
positions are presented. The 
comparison with previous results is discussed. For the kinematical analysis we 
used the parametrization for the LOSVD given in van der Marel et al. (1993).}

\FullConference{Baryons in Dark Matter Halos\\
                 5-9 October 2004\\
                 Novigrad, Croatia}

\begin{document}

\section{Introduction}

Hickson Compact Groups (hereafter HCGs) are small groups of four to ten 
galaxies in a dense configuration on the sky {\cite{Hickson82}}. For a long 
time, their nature, formation and evolution have been matter of a lively 
discussion {\cite{Hickson82}}. Due to the high galaxy density, similar to that
in the centers of rich galaxy clusters, and very low velocity dispersion 
(typically 200 km\,s$^{-1}$, comparable to the internal stellar motions 
of the group galaxies), HCGs are the perfect environment to study galaxy-galaxy interactions, effects of the Intra-Group Medium (IGM) on the morphological 
transformation of the galaxies in the group, and also the role of the dark 
matter in these compact structures.

HCG 90 is a galaxy group with four bright galaxies: two late-type galaxies 
NGC~7172 (H90a) and NGC~7174 (H90d), and two early-type galaxies, NGC~7176 
(H90b) and NGC~7173 (H90c). Three of the galaxies are located in the core of HCG 90 and are strongly interacting. H90d is more isolated, but nevertheless shows a disturbed structure {\cite{White03}}. There have been detected further member galaxies of HCG 90. In a region of 1$^{\circ}.$5 $\times$ 1$^{\circ}.$5, 19 fainter group members share the radial velocity of the bright members 
($\approx 2600\,{\rm km\,s^{-1}}$) and have a velocity dispersion of $\approx 190 {\rm km \, s^{-1}}$ {\cite{White03}}.

Another interesting feature in HCG 90 is the common stellar envelope around 
the three central galaxies. In a recent work by White et al. (\cite{White03}) 
the distribution of diffuse light in HCG 90 has been studied with some 
surprising results. The diffuse light appears to have a uniform red color which
is consistent with an old stellar population, typically found in elliptical 
galaxies {\cite{White03}}. Moreover, X-ray studies of HCG 90 show that the 
emission is not centered on any galaxy (which also has been observed in other 
dynamically young systems). The temperature of the gas is too low ($kT= 0.67 
\pm 0.03$ keV) to be consistent with the heating expected in a deep cluster 
potential {\cite{White03}}. It may be an indication that the amount of dark 
matter in HCGs is lower than in other structures like loose groups or galaxy 
clusters.

The questions that naturally arise are: Is HCG 90 a dynamically young system or an old one?
Will the galaxies merge within a short time?

To answer these questions and determine the evolutionary status of HCG 90 it is necessary to study the IGM/ISM  of HCG 90 as a tracer of interaction {\cite{Plana98}}. Unfortunately, HCG 90 is quite poor in HI 
(Verdes-Montenegro, priv. comm.). Only H90a has been detected in HI. And other 
wavelengths still remain to be explored (UV for example) {\cite{Plana98}}. 
Another possibility to study the interaction is the kinematical analysis of the
stellar populations {\cite{Bonfa99}}. However, there still exists controversy when 
this method is used, and HCG 90 is a perfect example for this. Whilst Longo et 
al. {\cite{Longo94}} claim that there is no evidence for interaction between 
H90b and H90d based on stellar kinematical features, Plana et al. 
{\cite{Plana98}}, using $H{\alpha}$ emission, found out that a possible 
interaction is taking place between the two galaxies. A possible explanation 
for this discrepancy is that the gas reacts faster to the interaction than the 
stellar component.         

\section{Observations and Data Reduction}

Longslit spectra were obtained in May 2002 with the VLT at Paranal (ESO). The 
spectra have a resoluution of 2-3\AA /pixel. The wavelength range covered is 
4600 $\le$ $\lambda$ $\le$ 5940 \AA. The slit lenght is 7.67 arcmin and the 
width 1 arcsec. The two slits connect the galaxy centers in the core of HCG 90.
The exposure time was 1800 seconds. The data reduction was performed with the
IRAF packages TWODSPEC and ONEDSPEC. The slit-exposures were bias subtracted, 
corrected for cosmic rays, and flat-fielded. The spectra were then corrected for their spatial distortion and illumination, wavelength calibrated and finally
log-rebinned for the kinematical analysis. The velocity resolution is about 
$50 \,{\rm km}\,/{\rm s}$.

\begin{figure}
\centering
\includegraphics[height=5.3cm,bbllx=3mm,bblly=65mm,bburx=198mm,bbury=208mm]{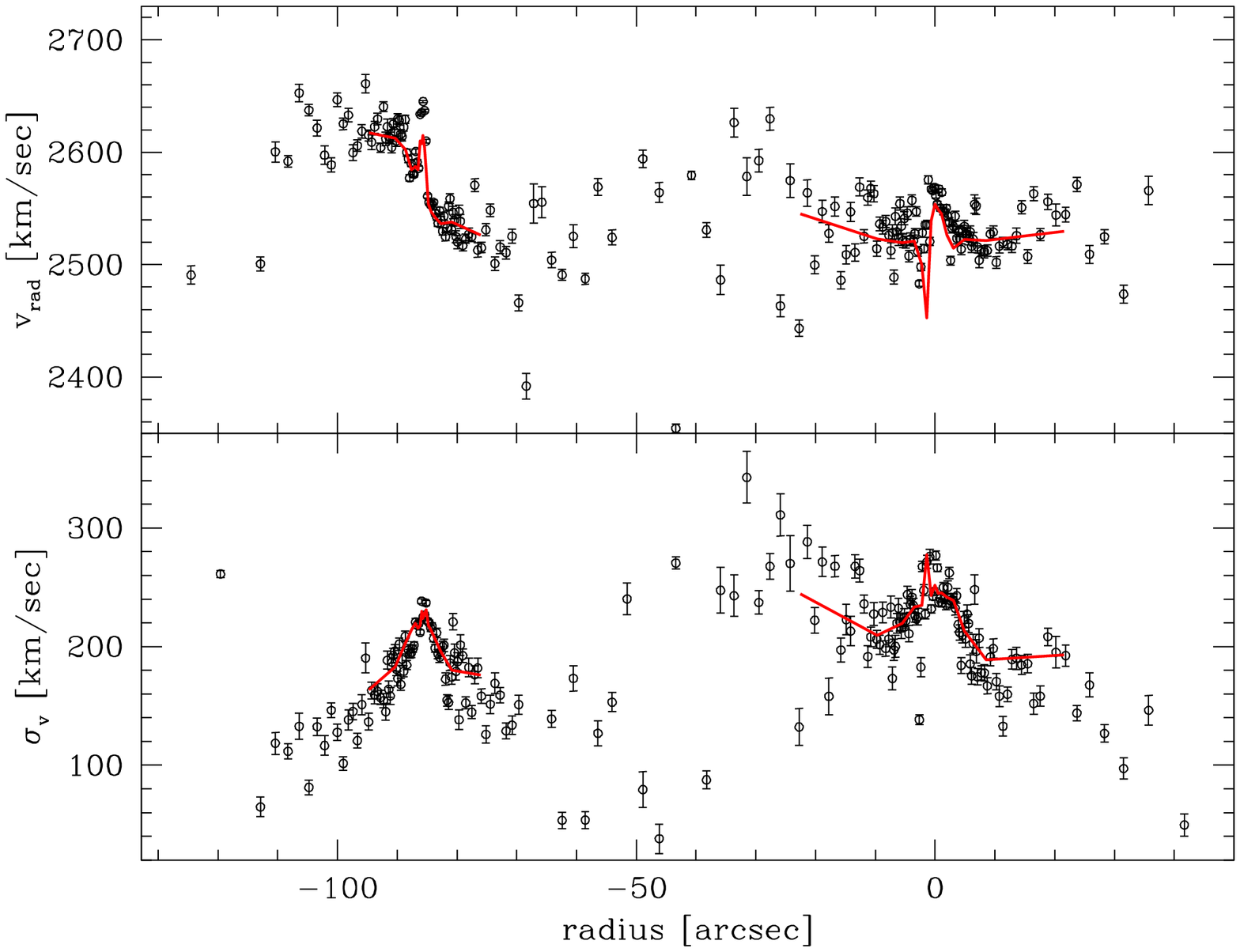}
\includegraphics[height=5.3cm,bbllx=3mm,bblly=65mm,bburx=198mm,bbury=208mm]{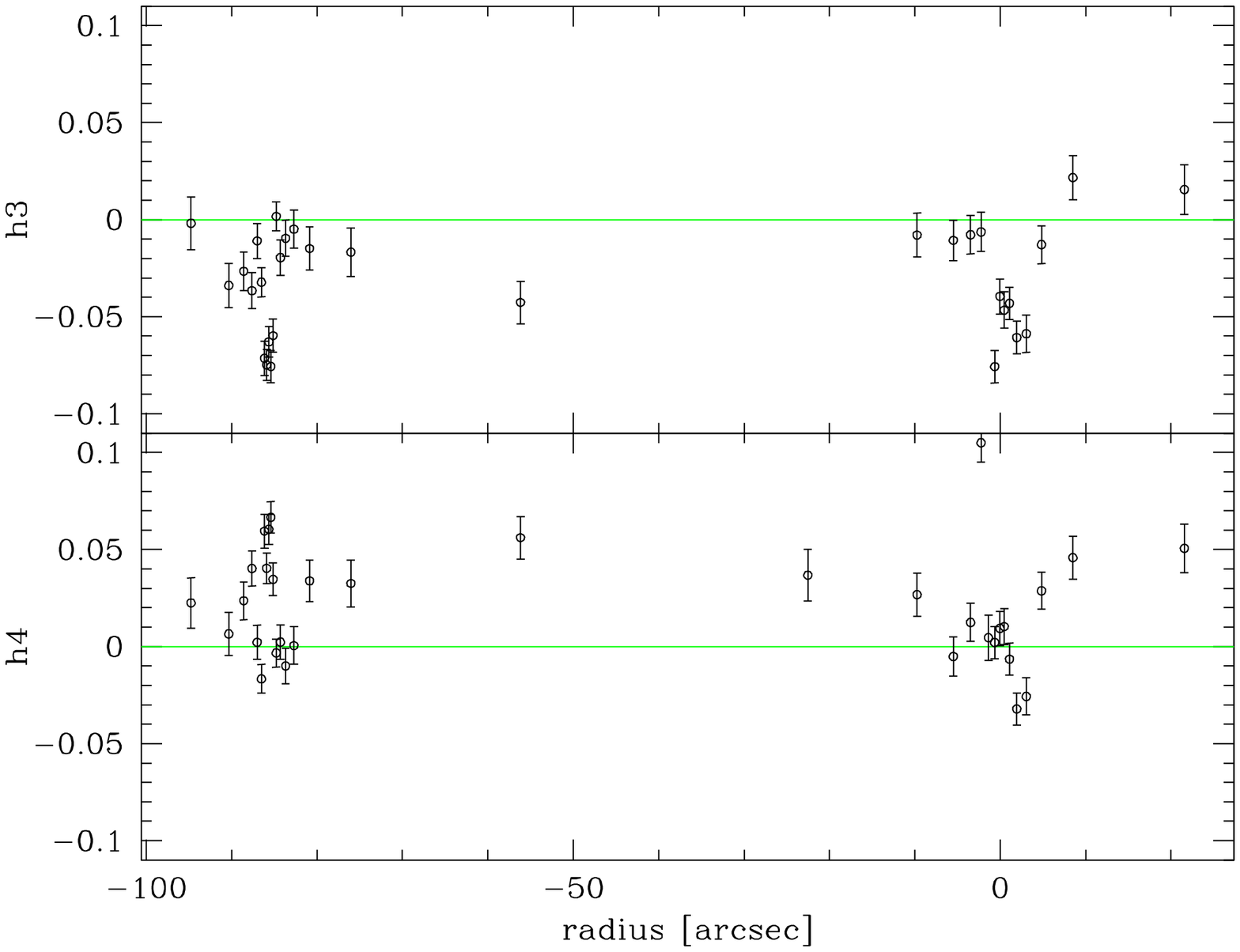}
\caption{\label {slit1} {\bf Left:} Top panel: Rotation curves of NGC
7176 and NGC 7173 at P.A.=132$^\circ$, for a S/N=50 (dots) and S/N=100 (solid curve) ; Lower panel: Velocity dispersion 
profile
profiles. {\bf Right} Hermite moments for binned spectra, with S/N=100.
}
\end{figure}
\section{Stellar Kinematics and velocity dispersion profiles}

The distribution of stellar velocities in the line of sight (hereafter LOSVD) 
was parametrized as a fourth-order Gauss-Hermite series {\cite{VdM94}}. These
contain basically the same parameters as the traditional Gaussian distribution,
but have two new coefficients, $h_3$ and $h_4$, that quantify the asymmetric 
and symmetric deviations from a Gaussian profile. For some mathematical 
properties and physical interpretation of the new terms see
{\cite{VdM94}}{\cite{Hau98}}. Contrary to isolated galaxies, few work has been 
done on interacting systems, and as we mentioned above, it is not an easy task 
to interpret kinematical signatures of stars for galaxies in interaction
{\cite{ca03}}.

For the kinematical analysis we used the software developed by van der Marel et
al. ({\cite{VdM94}}). The spectrum of a F7V star (HD 193901) was taken as 
stellar template. The fitting procedure was done in pixel space. The stellar
spectrum was convolved with Gaussian distributions of different mean velocities
and dispersions, and the residuals were minimized in the $\chi^2$ sense to
recover optimal values. For a complete description of the code see 
{\cite{Hau98}}.

\begin{figure}
\centering
\includegraphics[height=5.3cm,bbllx=3mm,bblly=65mm,bburx=198mm,bbury=208mm]{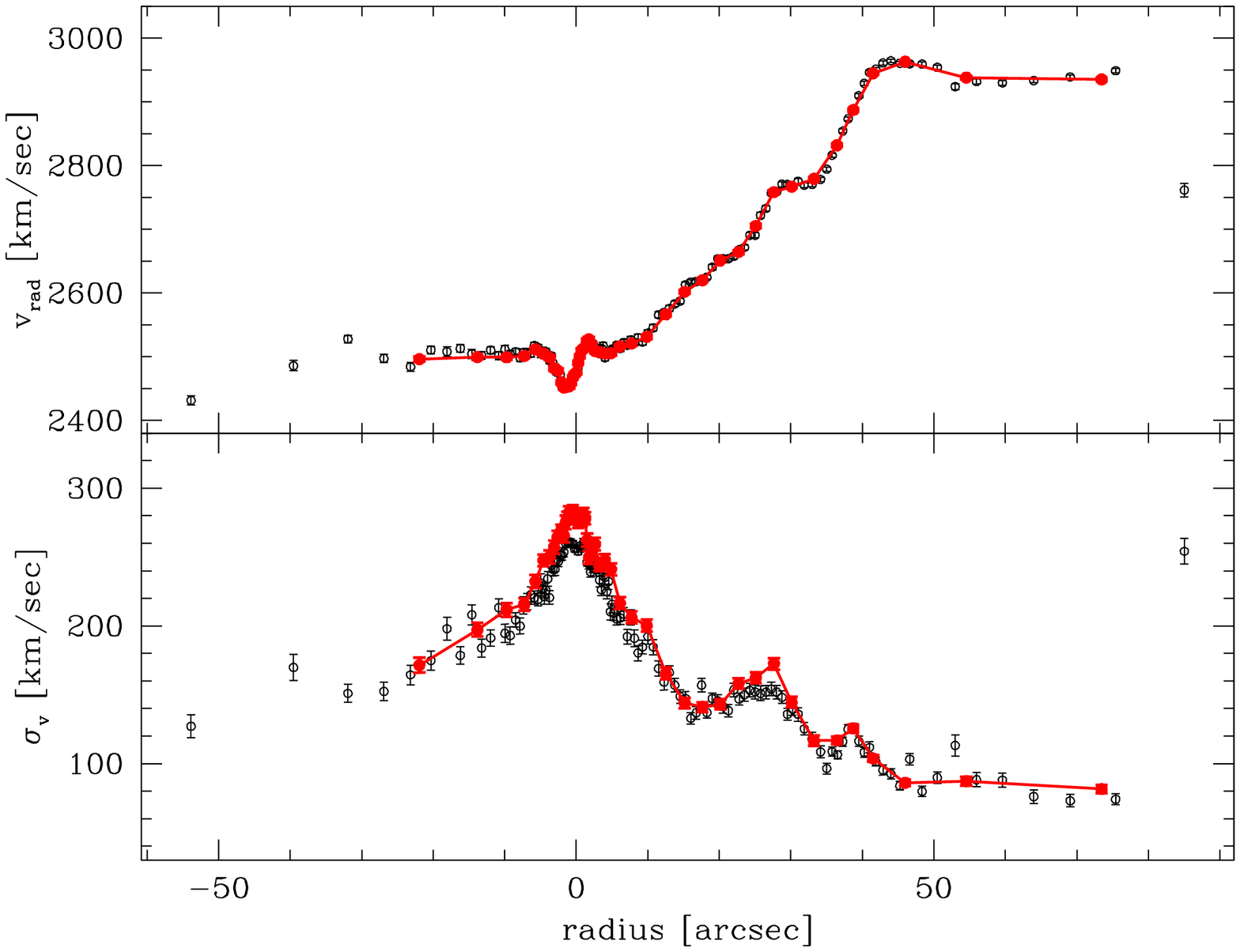}
\includegraphics[height=5.3cm,bbllx=3mm,bblly=65mm,bburx=198mm,bbury=208mm]{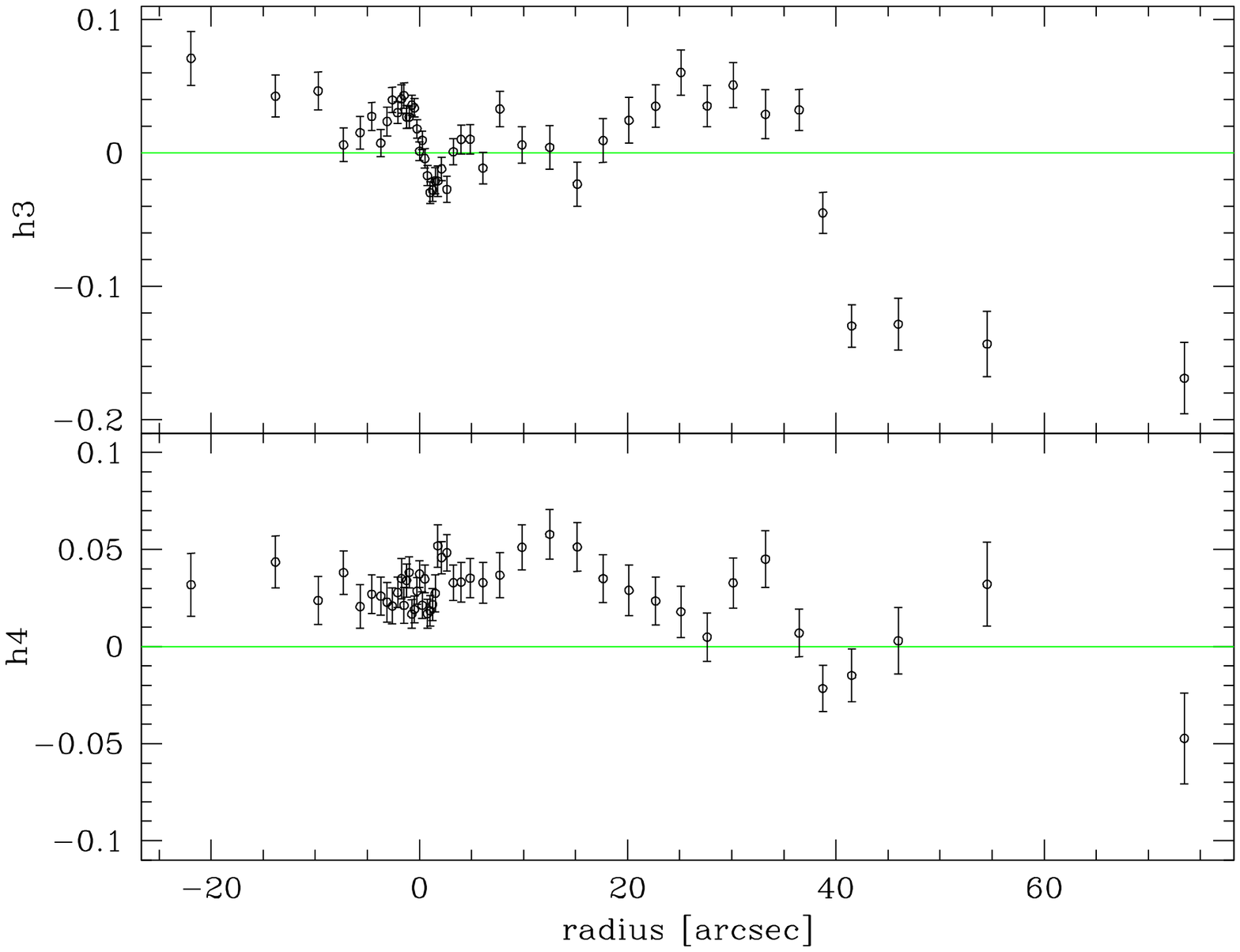}
\caption{\label {slit2}{\bf Left:} Top panel: Rotation curves of NGC 7176
(left) and NGC 7174 (right) at P.A.=72$^\circ$, Lower panel: Velocity
dispersion profile with the same convention. The S/N values as in Figure 1. {\bf Right:} Hermite 
moments for
binned spectra, with S/N=100.
}
\end{figure}

The results for the two longslit spectra are shown in the Figs.~{\ref{slit1}}
and {\ref{slit2}}. The first figure shows the kinematics for the slit 
connecting the two early-type galaxies H90b (left) and H90c (right). Both galaxies show signs of rotation. The velocity dispersion between the two galaxies scatters towards high values, probably originating from a highly disturbed velocity field.
The second figure shows the kinematics between H90b and H90d (the late-type spiral). The velocity field smoothly connects both galaxies with each other. Still one can not say wether this is due to interaction or rather is a superposition of the rotation curves of both galaxies. 
The results from both slits confirm the earlier results of Longo et al
{\cite{Longo94}}, but with quite a larger significance due to the much
higher signal-to-noise of our spectra. 

\section{Some Conclusions}
We have studied the stellar kinematics of HCG 90 using the Gauss-Hermite 
parametrization. Our results extend previos works to larger galactocentric radii. For the further analysis,  the rotation curves and the velocity dispersion profiles will be used together with analysis of the ionized gas to disentangle the interaction history of the system. In order to improve the significance of the 
the Gauss-Hermite moments, we have to minimize the mismatch between the galaxy  and the template spectra. However, already our preliminary result shows that there is a definite tendency in the behaviour of the moments (see right panels in Figs. 1+2). The physical 
interpretation of these findings has to be elaborated. It implies that new physics still remain to be discovered in interacting systems. The high quality of the velocity dispersion profiles will allow us to study the dark matter problem in HCG 90.

\end{document}